\documentclass[12pt]{article}
\usepackage[dvips]{graphicx}

\newlength{\extraspace}
\newlength{\extraspaces}
\newcommand{\be}{\begin{equation}
\addtolength{\abovedisplayskip}{\extraspaces}
\addtolength{\belowdisplayskip}{\extraspaces}
\addtolength{\abovedisplayshortskip}{\extraspace}
\addtolength{\belowdisplayshortskip}{\extraspace}}
\newcommand{\ee}{\end{equation}}

\newcommand{\ba}{\begin{eqnarray}
\addtolength{\abovedisplayskip}{\extraspaces}
\addtolength{\belowdisplayskip}{\extraspaces}
\addtolength{\abovedisplayshortskip}{\extraspace}
\addtolength{\belowdisplayshortskip}{\extraspace}}
\newcommand{\ea}{\end{eqnarray}}

\title{Application of novel techniques for interferogram analysis to
laser-plasma femtosecond probing.}
\author{\bf{ P. \, Tomassini, A.\, Giulietti,  L.A.\, Gizzi, R.\, Numico} \\
Intense Laser Irradiation Laboratory - IFAM CNR\\
  Area della Ricerca di Pisa\\
  Via G. Moruzzi, 1 56124 Pisa (Italy)\\
   {\it E. mail: tomassini@ifam.pi.cnr.it}\\
\bf{M.\, Galimberti and D.\, Giulietti} \\ Intense Laser
Irradiation Laboratory - IFAM CNR\\
   Dip. di Fisica Universita' di Pisa  and I.N.F.M unita' di Pisa\\
\bf{M.\, Borghesi}\\
 Dep. Of Pure and Applied Physics, The Queen's University, Belfast (UK)\\}

\begin{document}
\maketitle

\begin{abstract}
Recently, two novel techniques for the extraction of the
phase-shift map (Tomassini {\it et.~al.}, Applied Optics {\bf 40}
35 (2001)) and the electronic density map estimation (Tomassini P.
and Giulietti A. , Optics Communication {\bf 199}, pp 143-148
(2001) ) have been proposed. In this paper we apply both methods
to a sample  laser-plasma interferogram obtained with femtoseconds
probe pulse, in an experimental setup devoted to laser particle
acceleration studies.

\end{abstract}
\section{Introduction}
The phase-shift extraction from interferogram images is generally
performed via a straightforward method based on Fast Fourier
Transforms (FFT) (Takeda, 1982). This method is fast and generally
very effective, but it can fail in producing accurate phase-shift
maps in facing with low quality fringe structures. Unfortunately,
the number of physical processes that can degrade the fringe
visibility is very large. Even if the effects of some of these
sources can be strongly reduced with refinement of the
interferometry techniques (e.g. in the case of uncorrelated noise
in the image, non uniform illumination...) or a reduction of the
probe duration (in the case of fringe smearing due to the plasma
evolution), the effect of the deviation of the probe light due to
strong electronic density gradients cannot be eliminated in the
interferometer acquisition step. As a result, standard FFT
technique hardly work properly when applied to interferograms of
very steep density gradients. Recently (Tomassini {\it et.~al.},
2001) a new method to extract the phase-shift map from
interferograms has been proposed. The new method IACRE
("Interferogram Analysis via Continuous wavelet transform Ridge
Extraction") takes advantage of the combined spatial and spectral
resolution of the Continuous Wavelet Transforms (CWT)
(Holschneider, 1995) to identify the fringe structures and results
more flexible, accurate and robust than standard FFT based one.

Once the phase-shift map has been obtained, one has to face with
Abel Inversion in order to retrieve the $2D$ electronic density
map. The Abel Inversion method is based on the strong assumption
that the full $3D$ electronic density is axisymmetric along an
axis parallel to the interferometry plane. This assumption is
often poorly verified and if we  force in using Abel Inversion we
can introduce large errors in the density map, especially near the
best estimated symmetry axis. In a recent procedure (Tomassini and
Giulietti, 2001),  Abel Inversion has been generalized and can be
applied to density distributions with a moderate asymmetry. As a
corollary, this generalized Abel Inversion provides a first step
in $3D$ density map extraction.

In this paper we show the preliminary results of the analysis of
interferograms obtained with femtosecond laser pulses probing
plasmas produced by the exploding foils technique. The set of
interferograms has been produced in order to characterize the
plasma  before,  during and after the passage of a ultrastrong
laser pulse, in an experimental setup devoted to the study of the
production of multi-MeV electrons generated by laser-plasma
interactions (Giulietti {\it et.~al.}, 2001).

\section{Interferometry setup}
The experiment has been performed on the Salle Jaune of the
Laboratoire d'Optique Appliquee with a laser pulse of wavelength
$0.82 \mu m$  delivering $1 Joule$ in $35 fs$. A portion of the
pulse has been doubled in frequency with a KDP cristal (probe
pulse) and the remaining portion (main pulse) has been focused
with an off-axis parabola on the target, a thin foil of FORMVAR
(see Fig. \ref{fig:setup}).


The nanosecond Amplified Spontaneous Emission (ASE) which precedes
the main pulse makes the foil to explode in the focal spot. As a
consequence, the main pulse interacts with a preformed plasma of
peak density well below the critical density and of scalelength of
few tens of $\mu m$'s.

The interferometry  pulse probes the plasma in a direction
parallel to the plastic foil, i.e. perpendicular to the plasma
expansion axis which is also roughly  the plasma symmetry  axis. A
modified Normarski interferometrt setup (Benattar {\it et al.},
1979) is   used to generate the interferometry images on a CCD
camera (see Figg. \ref{fig:setup} and \ref{fig:interfsetup}).


\section{The phase-shift extraction with the IACRE method}
A detailed description of the IACRE method and an accurate
comparison between IACRE  and the standard FFT-based methods
performances has been published elsewere (Tomassini, Giulietti and
Gizzi, 2000; Tomassini {\it et.~al.}, 2001). In this section we
will just sketch the main steps of the novel method.

Consider the interferogram of Fig. \ref{fig:1601-3}. It has been
produced $10 ps$ after the interaction of  a preformed plasma with
a $35 fs $ main pulse focused on it with an intensity of $10^{20}
W/cm^2$.


 The laser radiation came from the right hand side and the plasma is
approximately symmetric along the horizontal  $x$ coordinate. The
two black arrows refer to the positions of the two images of the
target, while the boxes show the two analyzed regions, one on the
side of the incoming laser radiation (the right one) and one on
the rear side.

To apply the IACRE method to the each  sub-image (left and right
boxes), we extract the phase shift $\delta\phi(z,x)$  as follow.
For each $z$ build a sequence $s_z(x)$ by taking the line-out of
the sub-image. Next:
\begin{itemize}
\item{} compute the Continuous Wavelet Transform coefficients $C_s(a,b)$
of $s_z(x)$, being $a$ and $b$ the "voice" and "time" parameters,
respectively;
\item{} detect the  "Ridge" ${\cal R}(C_s)$ of the $C_s$ map,
recording the phase $\phi(x)$ of the signal rebuilted with ${\cal
R}(C_s)$;
\item{} estimate the wavevector $k_p$ of the unperturbed fringes and, finally,
compute the phase-shift $\delta\phi(z,x)$ as
\be
\label{deltaphi} \delta\phi(z,x) = \phi(z,x)-k_p x  \, .\ee
\end{itemize}

The analysis of the test interferogram is complicated by the
presence of fringe curvature also in absence of the plasma. This
problem has been solved by acquiring a "Mask" interferogram
(without plasma) before each shot. A phase-shift of the "Mask"
$\delta\phi^{Mask}$ is estimated and the corrected phase-shift due
to the electronic plasma density $\delta\phi^{Plasma}$ is obtained
simply as \be \delta\phi^{Plasma} = \delta\phi-\delta\phi^{Mask}\,
. \ee \noindent The results for the left and right boxes
sub-images are reported in Fig. \ref{fig:16left}.


The two phase-shift maps of the front (right box) and rear (left
box) sides of the plasma have quite similar structures, with
strong evidence of a hole in the electronic density near the
target, where steep density gradients occur. These gradients are
responsible of the reduction of fringe visibility  at the edges of
the hole.

\section{Non axisymmetric Abel Inversion}
We now proceed in the estimation of the electronic density map
$n_e$. As it is clear in Fig. \ref{fig:16left}, the phase-shift
maps are not really mirror symmetric so we expect that the
computation of $n_e$ via standard Abel Inversion  could introduce
relevant errors. We then apply the Generalized Abel Inversion
(Tomassini and Giulietti, 2001) to the corrected phase-shift maps
(left and right boxes) in order to minimize inversion errors.  As
in the standard Abel Inversion, the position $z_0$ of the best
symmetry axis must be determined by maximizing, for example, the
cross-correlation between the two half maps \ba
\delta\phi^+(\zeta,x) & = &\delta\phi^{Plasma}(z-z_0,x)\, \,
z>z_0\, ,\nonumber\\
 \delta\phi^-(\zeta,x) & = &\delta\phi^{Plasma}(z_0-z,x)\, \,
z<z_0\, ; \ea next two $2D$ maps $n_0(r,x)$ and $n_1(r,x)$ are
numerically computed with the integrals:
 \ba \label{GenAbel} n_0(r,x) &=& -n_c {\lambda_p\over
\pi^2} \int_r^{\infty}d\zeta {1\over \sqrt{\zeta^2-r^2}}{\partial
\over\partial \zeta} \delta\phi_s(\zeta,x) \nonumber\\
 n_1(r,x) &=& -n_c {\lambda_p\over
\pi^2} r \int_r^{\infty}d\zeta {1\over
\sqrt{\zeta^2-r^2}}{\partial \over\partial \zeta}
\left({\delta\phi_a(\zeta,x)\over \zeta}\right)\, , \ea \noindent
where $n_c$ is the critical density for the probe wavelength
$\lambda_p$ and $\delta\phi_s$, $\delta\phi_a$ are the symmetrized
and anti-symmetrized half maps: $$\delta\phi_s \equiv {1\over
2}\left(\delta\phi^+(\zeta)+\delta\phi^-(\zeta)\right)\, , \,\,
\delta\phi_a \equiv {1\over
2}\left(\delta\phi^+(\zeta)-\delta\phi^-(\zeta)\right)\, . $$

The $3D$ electron density map can now be built-up as follow.
First, the map is mirror-symmetric along the $x-z$ plane. This
assumption is necessary because in the process of formation of the
interferogram an integration along the $y$ direction is made.
Consider then one half space (let's say the one with $y$ positive)
and, for each $x$, identity a point in the $y-z$ plane by using
polar coordinates $(r,\theta)$. Finally, the three dimensional map
\be\label{poly} n(r,\theta,x) = n_0(r,x)+n_1(r,x)\, cos(\theta)
\ee \noindent represents the best estimation of the electronic
density map obtained with the Generalized Abel Inversion. In Fig.
\ref{fig:16dleftright} the projection in the $x-z$ plane of the
density map at both sides of the target is reproduced, while in
Fig. \ref{fig:densityslides} a sequence of slices of the $3D$
density map in the left box at constant $x$ is shown.



The projection of the density maps onto the $x-z$ plane (see Figg.
\ref{fig:16dleftright} and \ref{fig:densityslides}) confirms the
presence of a dramatic density depression (hole) along the laser
propagation path, near the original target  position.

\section{Comments}
We applied the IACRE and Generalized Abel Inversion to a sample
interferogram, obtained just few $ps$ after the interaction of an
ultraintense femtosecond laser pulse with a preformed plasma from
an exploding thin foil. The density map evidences the creation of
a sharp electron density depression near the original target
position.

We  mention that the methods has been also applied to a series if
interferograms taken before, during and after the propagation of
the ultrarelativistic $35 fs$ pulse with the plasma.

The analysis is still in progress and preliminary results confirm
(Tomassini {\it et al.}, 2001) that the IACRE method is much more
robust and sensitive than standard FFT-based method, which
frequently failed in producing reasonably accurate phase-shift
maps. We then expect that  IACRE and Generalized Abel Inversion
methods will give a strong contribution in the comprehension of
such an extreme laser-plasma interaction regime.

\section{References}

Benattar, R., Popovics, M., Siegel, R., {\it Polarized light
interferometer for laser fusion studies}, Rev.Sci.Instrum.
{\bf50}, 1583 (1979)

{Holschneider, M. ; {\it Wavelet: An analysis tool}, Clarendon
Press -Oxford (1995).

Takeda, M., Ina, H.,  Kobayashi, S. {\it Fourier-transform method
of fringe-pattern analysis for computer-based topography and
interferometry},  J.Opt.Soc.Am. {\bf 72}, 156 (1982).

Tomassini, P., Giulietti, A.,  and Gizzi,L.~A. {\it Analyzing
laser-plasma interferograms with the continuous Wavelet
Transform}, IFAM-Note 2/2000, 15-11-2000, available at
http://xray.ifam.pi.cnr.it.

Tomassini, P., Borghesi, M., Galimberti, M., Giulietti,A.,
Giulietti, D.,  Willi, O., Gizzi, L.A. {\it Analyzing laser-plasma
interferograms with a Continuous Wavelet Transform Ridge
Extraction technique},  Applied Optics. {\bf 40} 35 (2001).

Tomassini, P. and Giulietti, A. {\it A generalization of Abel
Inversion to non axisymmetric density distribution}, Opt. Comm.
{\bf 199}, pp 143-148 (2001).

\section{Figure Captions}

Fig.~1 Interferometer setup: a modified Nomarski interferometer
has been used to generate digital interferometry images. A probe
pulse (a fraction of the second harmonic of the main pulse) probes
the plasma perpendicularly to the symmetry axis.

Fig.~2 A plastic foil explodes under the intense ASE radiation
coming along the $x$ axis. The plasma is approximately symmetrical
along the $x$ axis. The probe beam probes the plasma along the $y$
direction and generates an interferometry image on a CCD camera.

Fig.~3 Interferometry  pattern of a plasma obtained from $1\mu m$
thick foil. The probe pulse followed the main pulse by $10 ps$.
The laser pulse came from the right. The  (a) and  (b) boxes refer
to the two analyzed  regions of the plasma (left and right sides
of the foil target, respectively). The two images of the thin foil
are indicated by the black arrows.

Fig.~4 Corrected phase-shift of the left (a) and right (b) boxes
of Fig. 3.

Fig.~5 Projection of the density map in the left and right boxes
of the interferogram of Fig. \ref{fig:1601-3} onto the $x-z$
plane. A hole in the electronic density map is apparent.

Fig.~6 Slides of the $3D$ electron density map in the left box of
the interferogram of Fig. \ref{fig:1601-3}. A hole in the
electronic density is evident at distance from the target between
$x=15 \mu m$ and $x = 30 \mu m$.

\section*{Figures}
\newpage
\begin{figure}
\begin{center}
\includegraphics[angle=90,width=10cm]{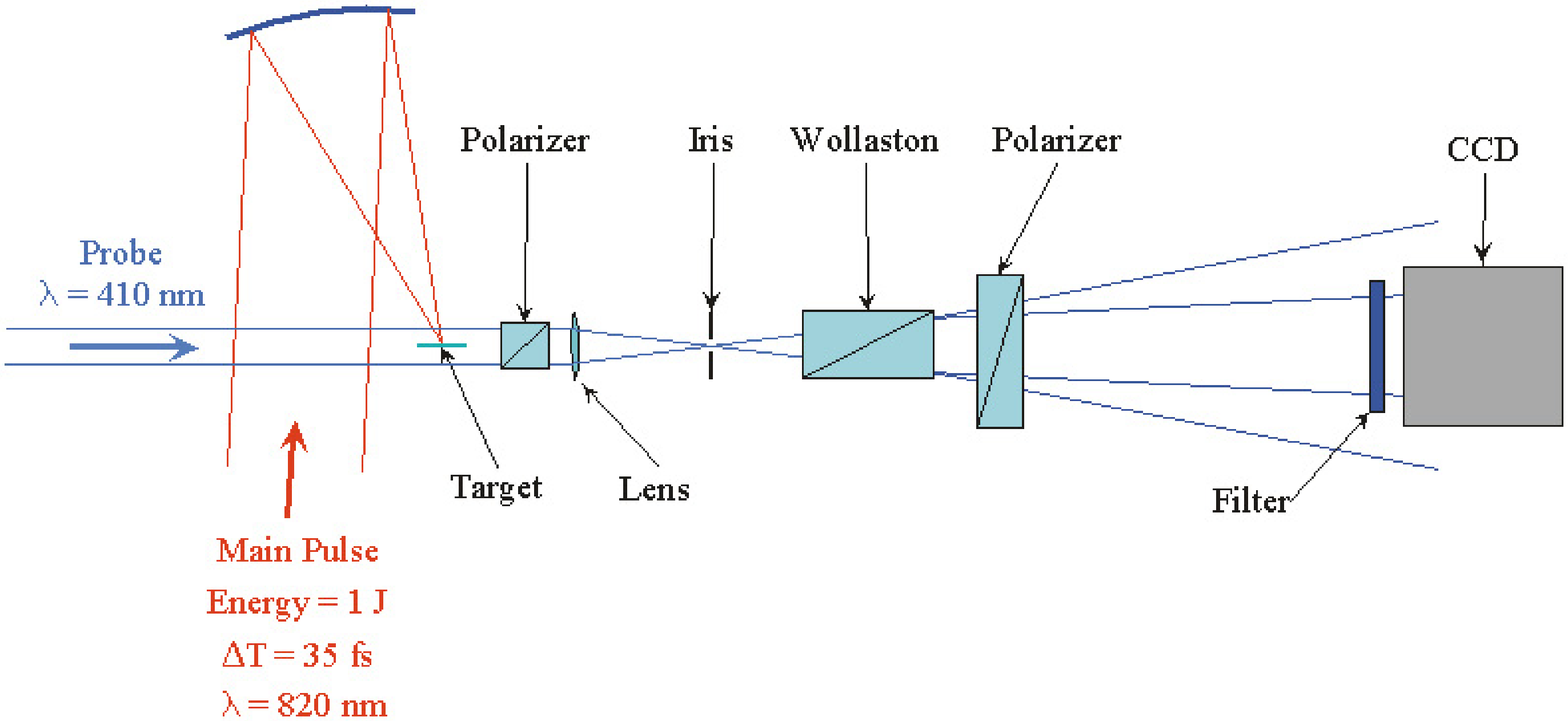}
\caption{}\label{fig:setup}
\end{center}
\end{figure}

\newpage
\begin{figure}
\begin{center}
\includegraphics[angle=90,width=12cm]{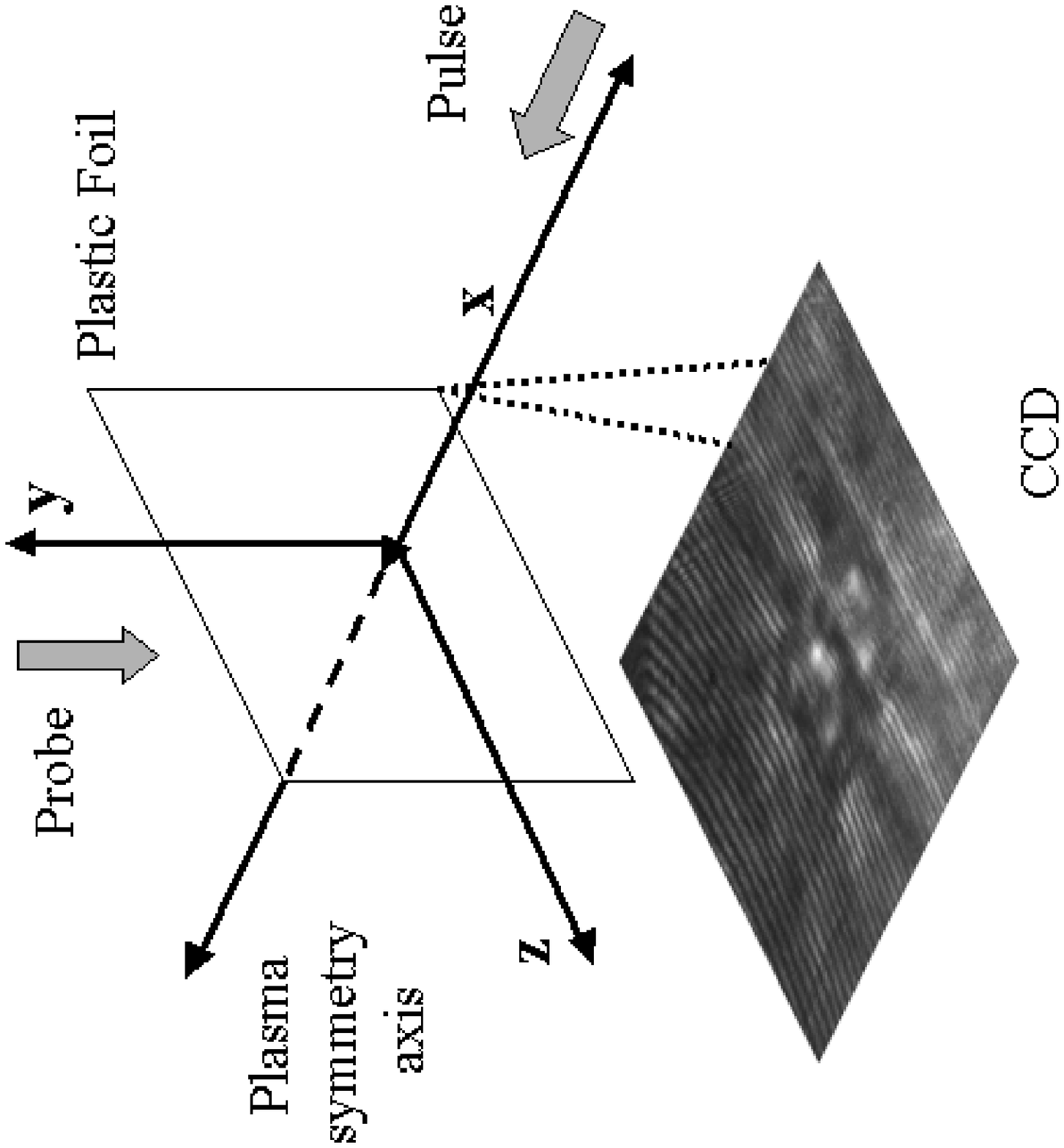}
\caption{}\label{fig:interfsetup}
\end{center}
\end{figure}

\newpage
\begin{figure}
\begin{center}
\includegraphics[angle=90,width=12cm]{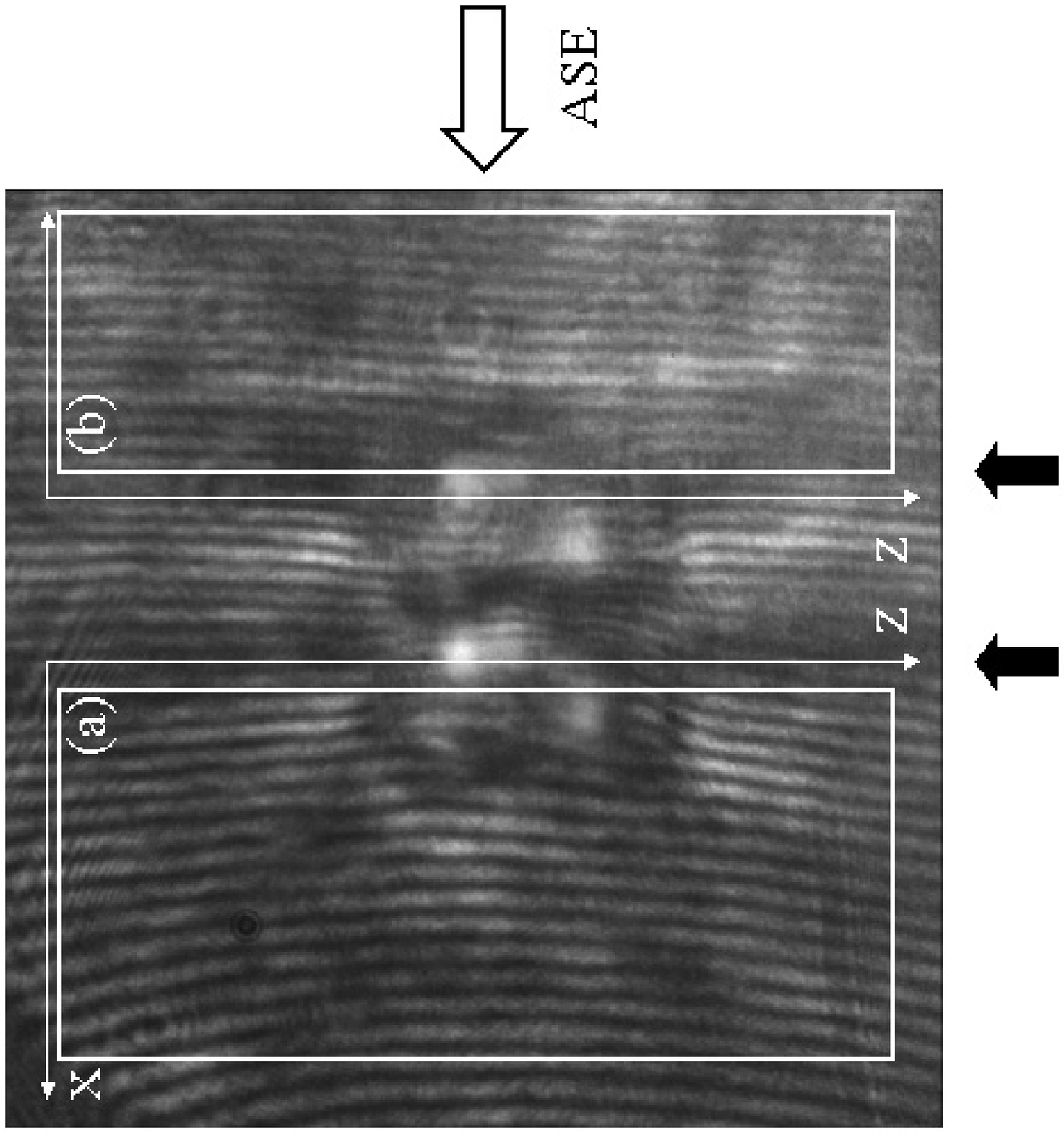}
\caption{}\label{fig:1601-3}
\end{center}
\end{figure}

\newpage
\begin{figure}
\begin{center}
\includegraphics[angle=90,width=9cm]{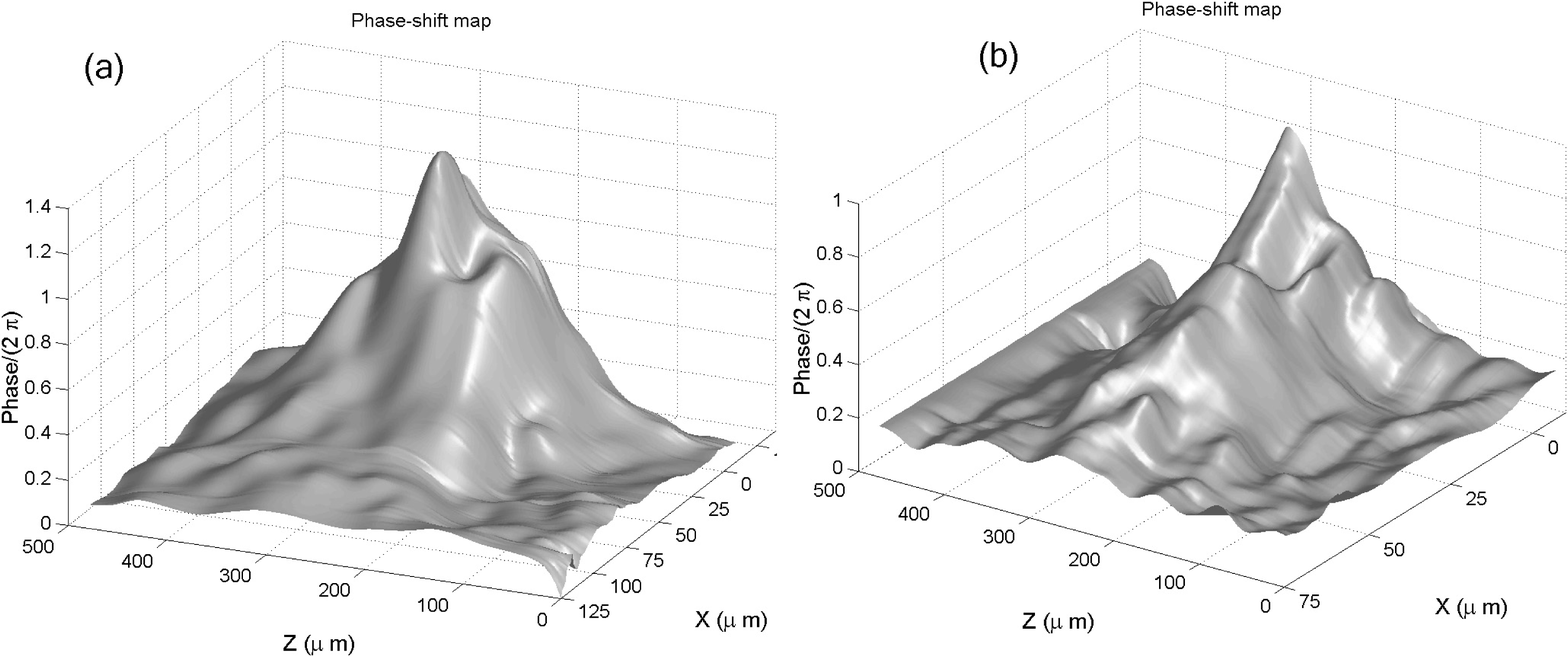}
\caption{}\label{fig:16left}
\end{center}
\end{figure}

\newpage
\begin{figure}
\begin{center}
\includegraphics[angle=90,width=10cm]{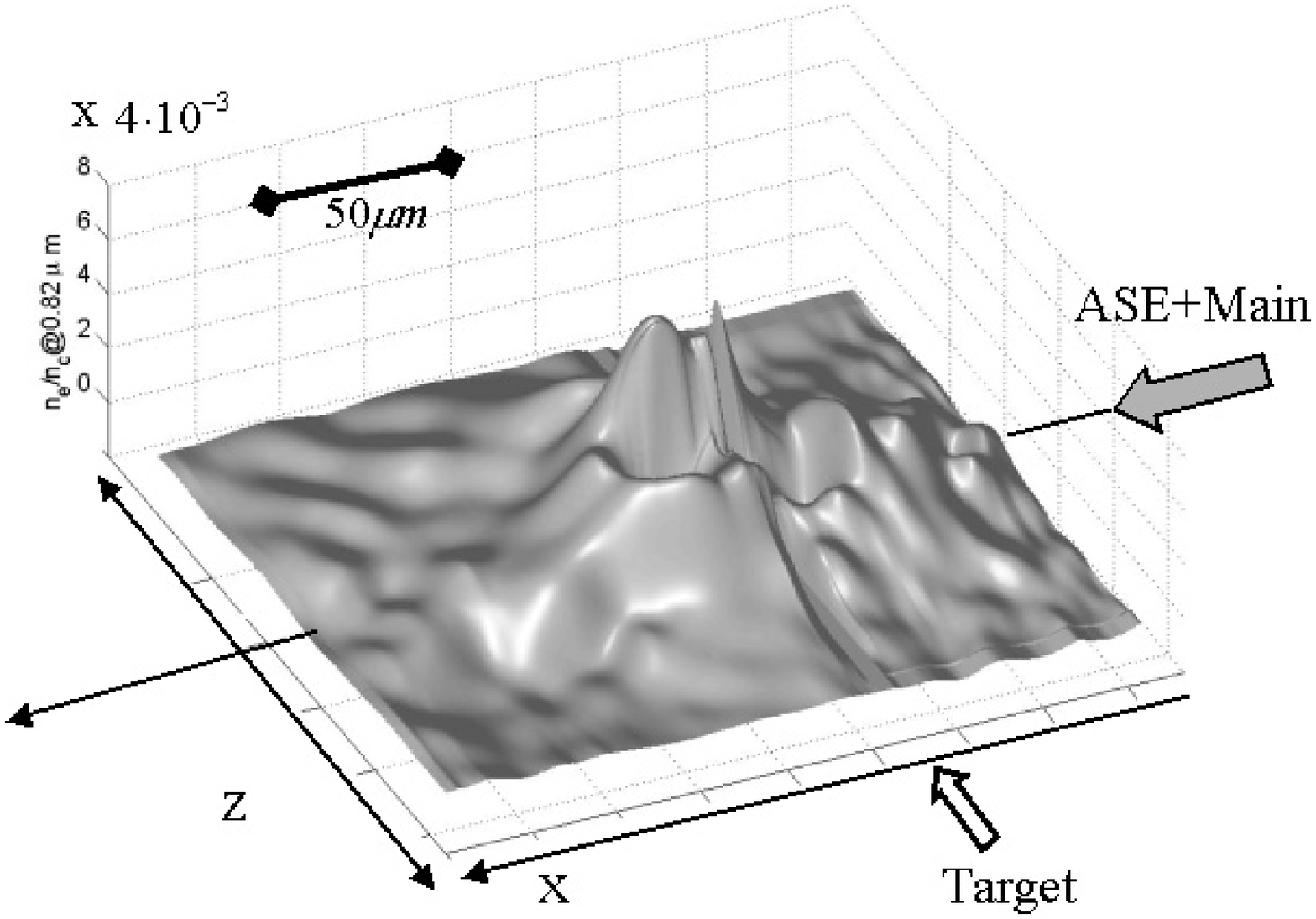}
\caption{}\label{fig:16dleftright}
\end{center}
\end{figure}

\newpage
\begin{figure}
\begin{center}
\includegraphics[angle=90,width=10cm]{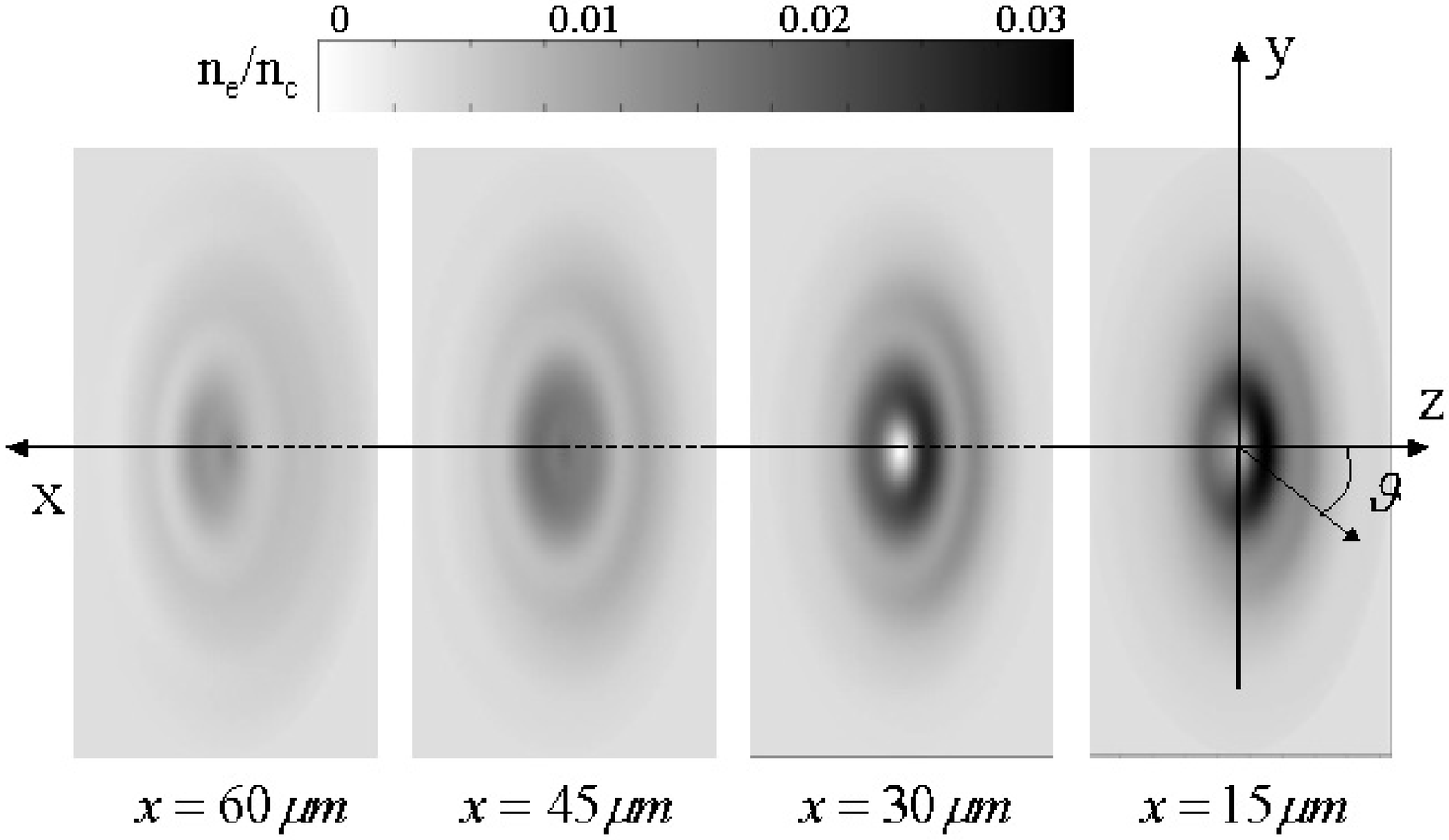}
\caption{}\label{fig:densityslides}
\end{center}
\end{figure}

\end{document}